\documentclass[9pt]{article}
\usepackage{spconf,amsmath,graphicx}
\usepackage[noadjust]{cite}

\title{Music Auto-tagging with Robust Music Representation Learned via 
Domain Adversarial Training}
\name{\parbox{0.7\linewidth}{\centering
    Haesun Joung\(^{1}\),
    Kyogu Lee\(^{1,2,3}\)}
}
\address{\(^1\)Department of Intelligence and Information, Seoul National University \\
\(^2\)Interdisciplinary Program in Artificial Intelligence, Seoul National University\\
\(^3\)Artificial Intelligence Institute, Seoul National University}

\begin{document}
\ninept
\maketitle
\begin{abstract}
Music auto-tagging is crucial for enhancing music discovery and recommendation. Existing models in Music Information Retrieval (MIR) struggle with real-world noise such as environmental and speech sounds in multimedia content. This study proposes a method inspired by speech-related tasks to enhance music auto-tagging performance in noisy settings. The approach integrates Domain Adversarial Training (DAT) into the music domain, enabling robust music representations that withstand noise. Unlike previous research, this approach involves an additional pretraining phase for the domain classifier, to avoid performance degradation in the subsequent phase. Adding various synthesized noisy music data improves the model's generalization across different noise levels. The proposed architecture demonstrates enhanced performance in music auto-tagging by effectively utilizing unlabeled noisy music data. Additional experiments with supplementary unlabeled data further improves the model's performance, underscoring its robust generalization capabilities and broad applicability.

\end{abstract}
\section{Introduction}
\label{sec:intro}

Music auto-tagging is the automated process of attaching relevant semantic labels such as genre, mood, or instrument to musical tracks, usually enabled by machine learning algorithms. This function is crucial for effective music information retrieval, personalization, and recommendation systems, predominantly in music-streaming platforms such as Spotify. These services rely heavily on clean, pure music tracks and utilize comprehensive metadata for each track to create a tailored and enriched user experience. This metadata, originating from clean musical sources, allows for precise alignment with individual user preferences.

\begin{figure}[tb]
\setlength{\belowcaptionskip}{-10pt}
    \begin{minipage}[b]{1.0\linewidth}
      \centering
      \centerline{\includegraphics[width=8.5cm]{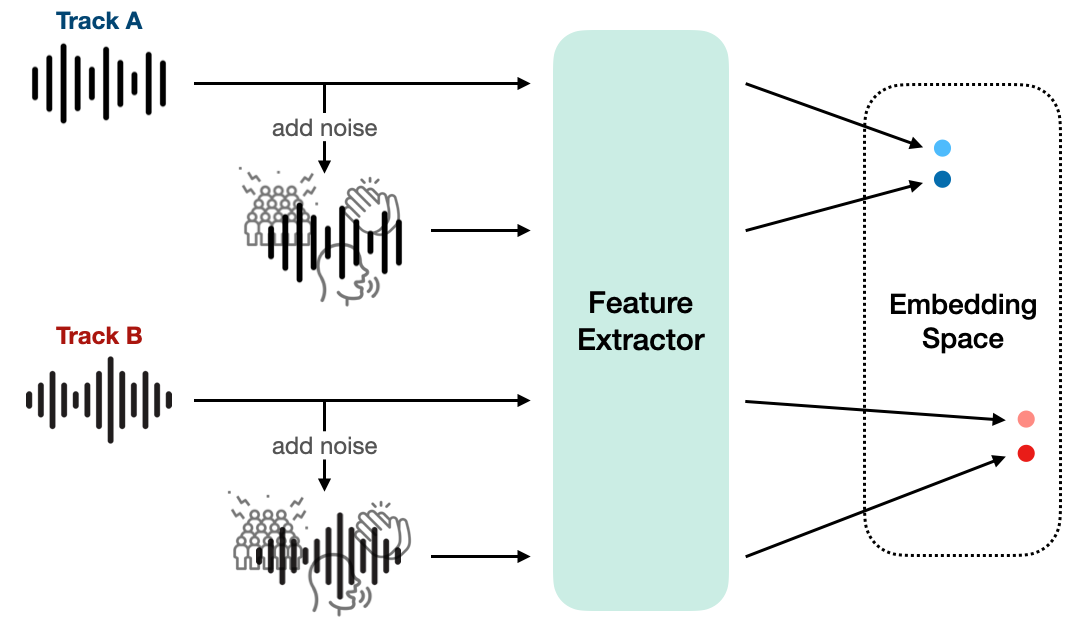}}
    \end{minipage}
    \caption{Feature extraction from clean and noisy music tracks in robust music representation learning. The extractor aims to produce closely positioned embeddings for the same track, regardless of audio quality.}
    \label{fig:FE_fig1}
\end{figure}

Furthermore, music auto-tagging is not only crucial for enhancing search capabilities and user accessibility in music streaming services, but also vital for catering to the specific needs of users who demand more personalized recommendations and detailed search options for music content in video-streaming platforms like YouTube \cite{choi2023towards}. Through the use of meaningful semantic tags, users can more effectively search for specific genres, artists, and moods, thereby enhancing the overall user accessibility and discoverability of music contents. However, the challenge is compounded on video-streaming services where music tracks are frequently mixed with real-world noises like crowd sounds and applause. This complicates the task for existing auto-tagging algorithms, which are predominantly trained on clean music tracks. Given the limited diversity of current tags and the sheer volume of diverse music-related content uploaded daily, there is a compelling need to advance auto-tagging techniques. Such improvements will not only make searching more efficient but also significantly contribute to delivering a more personalized and enriched user experience across various platforms.

As existing models have difficulty maintaining consistent feature extraction from both clean and noisy versions of the same track, we propose creating a dataset that includes both clean and noisy versions of identical music tracks. To enhance the robustness of music representation, we employ the technique of Domain Adversarial Training (DAT) \cite{ganin2016domain} which was effective in improving noise robustness of speech representation with performing downstream tasks such as Automatic Speech Recognition (ASR) or Speaker Identification (SID) \cite{huang2022improving}. This method is designed to condition the feature extractor to be indifferent to whether a music track is clean or noisy, effectively diminishing the distinctions between the clean and noise domains. It accomplishes this by closely aligning the embeddings of identical music tracks across both clean and noisy versions. As a result, this strategy allows the model to perform downstream tasks on noisy input representations as effectively as it would on clean representations. This refined framework is designed to facilitate model training for the recognition of musical elements within noisy environments, thereby improving overall model performance across a diverse range of auditory conditions.

\begin{figure*}[tb]
\setlength{\belowcaptionskip}{-7pt}
    \begin{minipage}[b]{1.0\linewidth}
      \centering
      \centerline{\includegraphics[width=15cm]{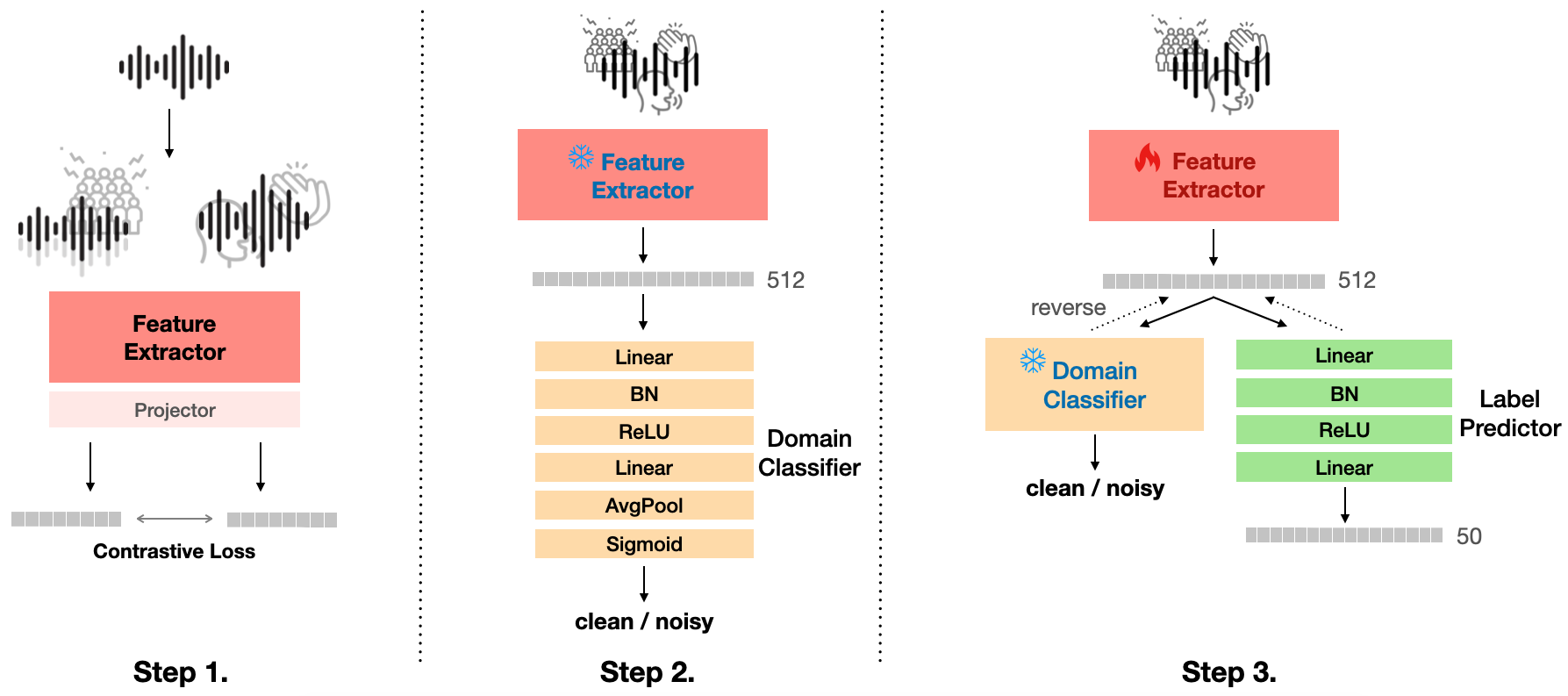}}
    \end{minipage}
    \caption{The proposed architecture and training process. Overall structure is composed of Feature Extractor (FE, pink), Domain Classifier (DC, yellow), and Label Predictor (LP, green). The training process is set to 3 steps for 1) pretraining FE, 2) pretraining DC, and 3) finetuning FE and training LP. In contrast, for both the baseline and oracle configurations, only the FE and LP are utilized, leading to a simplified two-step training process.}
    \vspace{-0.5cm}
    \label{fig:DAT_process}
\end{figure*}

\section{RELATED WORKS}

In the realm of \textbf{music auto-tagging}, Convolutional Neural Network (CNN)-based models \cite{krizhevsky2012imagenet} have demonstrated noteworthy performance, as evidenced by multiple studies \cite{choi2017convolutional, choi2016automatic, lee2017sample, pons2017end, kim2018sample}. \cite{choi2017convolutional} employed a hybrid model, integrating Recurrent Neural Networks (RNNs) with CNNs to more effectively capture temporal patterns in the data. Another study by \cite{choi2016automatic} deployed fully convolutional neural networks comprising multiple layers but without fully connected layers, thereby reducing the model's parameter count. While the majority of research in this area traditionally utilizes mel-spectrogram inputs, \cite{lee2017sample} diverged by using sample-level input without the mel-spectrogram conversion. Adding another layer of complexity, \cite{kim2018sample} incorporated a Squeeze-and-Excitation (SE) block into their sample-level input model, enhancing the extraction of representational features.

Nevertheless, given that music representations gain interpretability and significance when modeled sequentially, \cite{won2019toward} employed the Transformer architecture \cite{vaswani2017attention} as the backbone model for their study. Further, in subsequent work, \cite{won2021semi} adopted a semi-supervised approach in conjunction with the Transformer, highlighting the insufficiency of available data specifically for music auto-tagging. 
Furthermore, \cite{mccallum2022supervised, li2023mert, yuan2023marble}, collectively advance the field of music AI by exploring efficient pre-training strategies for audio understanding, \cite{li2023mert} introducing a novel self-supervised model (MERT) for nuanced music audio analysis, and \cite{yuan2023marble} establishing MARBLE, a comprehensive benchmark for evaluating music information retrieval systems.

The \textbf{Domain Adversarial Training} (DAT) \cite{ajakan2014domain} approach has been effective in making speech representation more robust, as shown in previous work \cite{huang2022improving}. In this setup, clean audio is used as the source domain, while different types of distorted audio make up the target domain. By applying the reversed gradient of the domain classifier's loss, the feature extractor can be tuned to lessen the difference caused by these distortions. This adjustment allows for better performance in downstream tasks using the label predictor. In this paper, we use the DAT approach from earlier research \cite{huang2022improving}, but with changes in domain settings and training steps, which we will discuss in subsequent sections.

\section{METHOD}

\subsection{Architecture}

In the proposed architecture, the model is composed of three primary components: Feature Extractor, Domain Classifier, and Label Predictor.

The \textbf{Feature Extractor} (FE) is first trained to extract general music embedding from the input audio, then finetuned to blur the distinction between clean and noisy input. In this paper, we employ CLMR \cite{spijkervet2021contrastive}, with SampleCNN \cite{lee2018samplecnn} serving as the encoder, whose backbone is SimCLR \cite{chen2020simple}. For FE, we exclusively utilize the encoder component of CLMR. The output embedding of the Feature Extractor then subsequently serves as the input for both the Domain Classifier and the Label Predictor.

The \textbf{Domain Classifier} (DC) is tasked with determining the origin of the embedding—whether it is derived from a clean or noisy audio source. The structure of the DC is based on the original DAT research \cite{ganin2016domain}. This module outputs a scalar that classifies whether the embedding originated from the clean source input or the noisy target input, which comprises simple fully-connected layers, accompanied by activation functions and batch normalization.

The \textbf{Label Predictor} (LP) focuses on the downstream task of music auto-tagging based on the provided embedding. Among the models in the proposed architecture, the LP stands out with its compact structure and minimal number of layers and parameters. This module takes the output from the FE as input and sequentially processes it through two fully-connected layers, with a ReLU activation function in between.

\begin{figure*}[tb]
\setlength{\belowcaptionskip}{-7pt}
    \begin{minipage}[b]{1.0\linewidth}
      \centering
      \centerline{\includegraphics[width=18cm]{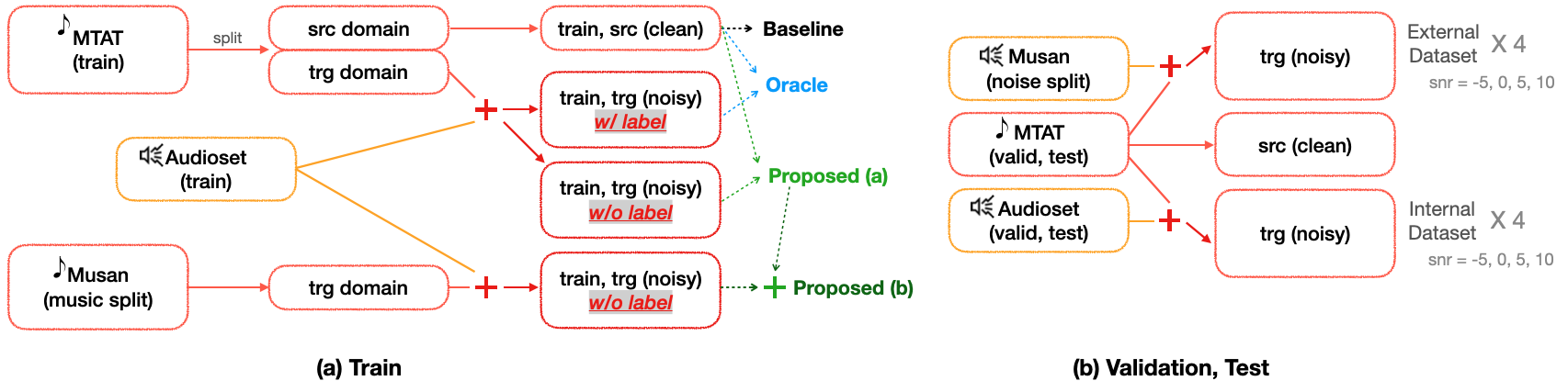}}
    \end{minipage}
    \caption{Proposed dataset configuration: The music dataset (red) utilizes MTAT \cite{law2009evaluation} and the music split from Musan \cite{snyder2015musan}. The real-world noise dataset (yellow) incorporates Audioset \cite{gemmeke2017audio} and the noise split from Musan Synthesized samples combining music and noise are designated as target domain data. During training, source (src) and target (trg) domain samples do not overlap. Note that the Musan noise dataset is exclusively employed for creating the test set, while it is not used in the formation of the validation set.}
    \label{fig:dataset}
\end{figure*}

\subsection{Training Process}
Our proposed training methodology incorporates elements from previous work \cite{huang2022improving}, but introduces an additional pretraining step for the DC, resulting in a three-step process in total.

The initial stage involves pretraining the FE. This phase allows the FE to gain a general understanding of both music representations and real-world noises, employing a contrastive loss function \cite{spijkervet2021contrastive, chen2020simple} for learning. In this step, both the encoder and the projector are trained, although only the encoder is used in subsequent steps. For an input audio $x_i$, the FE extracts the embedding of the audio $e_i$ as the result of $\text{FE}(x_i)$, which is a 512-dimensional vector.

\begin{equation*}
    \begin{aligned}
        (e_i, e_j) &= \text{FE}(x_i, x_j) \\
        (h_i, h_j) &= \mathrm{proj}(e_i, e_j) \\
        \mathcal{L}_{FE} &= \text{ContrastiveLoss}(h_i, h_j) \\
    \end{aligned}
\end{equation*}

The second stage focuses on the pretraining of the DC, with the parameters of the FE being frozen. This separate training step for DC deviates from the methods outlined in \cite{ganin2016domain} and \cite{huang2022improving}. We opted for this separation of training steps after observing a decline in performance when the Domain Classifier (DC) was trained alongside other components. Given an input embedding vector $e_i$, the DC performs binary classification to determine whether the input originates from clean or noisy musical audio.

\begin{equation*}
    \begin{aligned}
        \hat{d_i} &= \text{DC}(e_i)  \\
        \mathcal{L}_{DC} &= \text{BCELoss}(\hat{d_i}, d_i)
    \end{aligned}
\end{equation*}

The final stage is dedicated to the finetuning of the FE and the training of the LP. At this stage, the parameters of the DC are frozen, based on the assessment that it has achieved adequate binary classification performance. In contrast, LP is trained from scratch. For an input embedding vector $e_i$, the LP outputs a 50-dimensional vector, which corresponds to the classification of the 50 multi-tags in the MTAT dataset \cite{law2009evaluation}.
The combined loss of the LP and DC informs the fine-tuning of the FE, yielding the total loss function as described in the equation. Note that the gradient of the DC is negated which forces the FE to blur the distinction between clean and noisy domain. Also, $\mathcal{L}_\text{LP}^{\text{trg}}$ is not applicable, as tag labels for the target domain are assumed to be unavailable.
\vspace{-0.2cm}

        \begin{gather*}
            (e^{\text{src}}, e^{\text{trg}}) = \text{FE}(x^{\text{src}}, x^{\text{trg}}), \quad
            (\hat{d}^{\text{src}}, \hat{d}^{\text{trg}}) = \text{DC}(e^{\text{src}}, e^{\text{trg}}) \\
            \mathcal{L}_\text{LP} = \text{BCEWithLogitsLoss}( \hat{l}^{\text{src}}, l^{\text{src}} ), \quad \hat{l}^{\text{src}} = \text{LP}(e^{\text{src}}) \\
            \mathcal{L}_{\text{Total}} = \mathcal{L}_\text{LP}^{\text{src}} + \lambda \{ \mathcal{L}_\text{DC}^{\text{src}} + \mathcal{L}_\text{DC}^{\text{trg}} \}
        \end{gather*}

\section{DATASET}
In our experiments, we used the MTG-Jamendo dataset \cite{bogdanov2019mtg} for the pretraining of the FE and used the MagnaTagATune (MTAT) dataset \cite{law2009evaluation}  for music auto-tagging tasks. Regarding the size of the full Jamendo dataset, we selected a subset comprising audio files with both genre and mood/theme tags. For the real-world noise dataset, Audioset \cite{gemmeke2017audio} is used after filtering to exclude any data containing music-related tags, such as those denoting musical notes, for example, \textit{`bell'} or \textit{`ding'}. Additionally, the Musan dataset \cite{snyder2015musan} is used as an extra dataset which is provided in music, speech, and noise splits. We employed the music and noise splits for training and testing phases, respectively.

\subsection{Data Configuration}
For the pretraining of the FE, we utilized the Jamendo and Audioset datasets. The audio samples from Jamendo are subjected to random augmentations, such as pitch shifting and the application of filters, following the methodology outlined by \cite{spijkervet2021contrastive}. Additionally, to improve the model's generalization capabilities with respect to noisy musical audio, we synthesized random samples from Audioset with the Jamendo audio samples.

Regarding the subsequent steps in our training process, we employed the MTAT and Audioset datasets under diverse experimental configurations. In the baseline setting, we assumed that only clean audio samples with corresponding tags are accessible, which aligns with the existing frameworks for auto-tagging. For the oracle setup, we assumed the availability of tags for both clean and noisy audio samples, a condition that is not feasible in real-world scenarios. In the proposed experimental setting, we made use of clean audio samples with tags for the source domain, and synthesized noisy samples without tags for the target domain (\textit{proposed (a)}). Lastly, to demonstrate the capacity for further model training and generalization with noisy musical samples, we also utilized additional noisy samples synthesized from the Musan music split and Audioset, excluding any associated tags (\textit{proposed (b)}). This final setting serves to underscore the efficacy of the proposed architecture in accommodating real-world audio conditions.

For the validation and test phase, we did not split the MTAT dataset but fully and repeatedly used the validation and test dataset in five different conditions. First we used music audio samples without any noise added for clean source domain. From second to the last condition, we added noise but in different sound-to-noise ratio (SNR) conditions from -5 to 10.

    \begin{table*} [!hbt]
    \centerline{{\includegraphics[width=18cm]{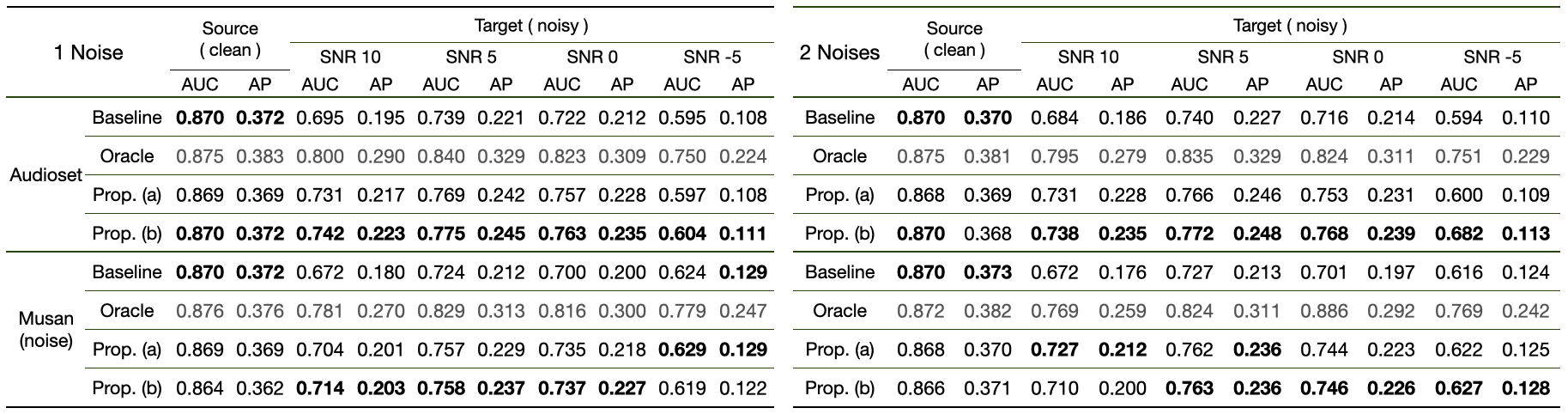}}}
    \caption{The test AUC and AP metrics for the baseline, oracle, and proposed configurations (a) and (b), evaluated with the inclusion of either 1 or 2 noises in the synthesized noisy music data.}
    \vspace{-0.4cm}
    \label{fig:noise1}
    \end{table*}

    \begin{table} [!hbt]
    \centerline{{\includegraphics[width=9cm]{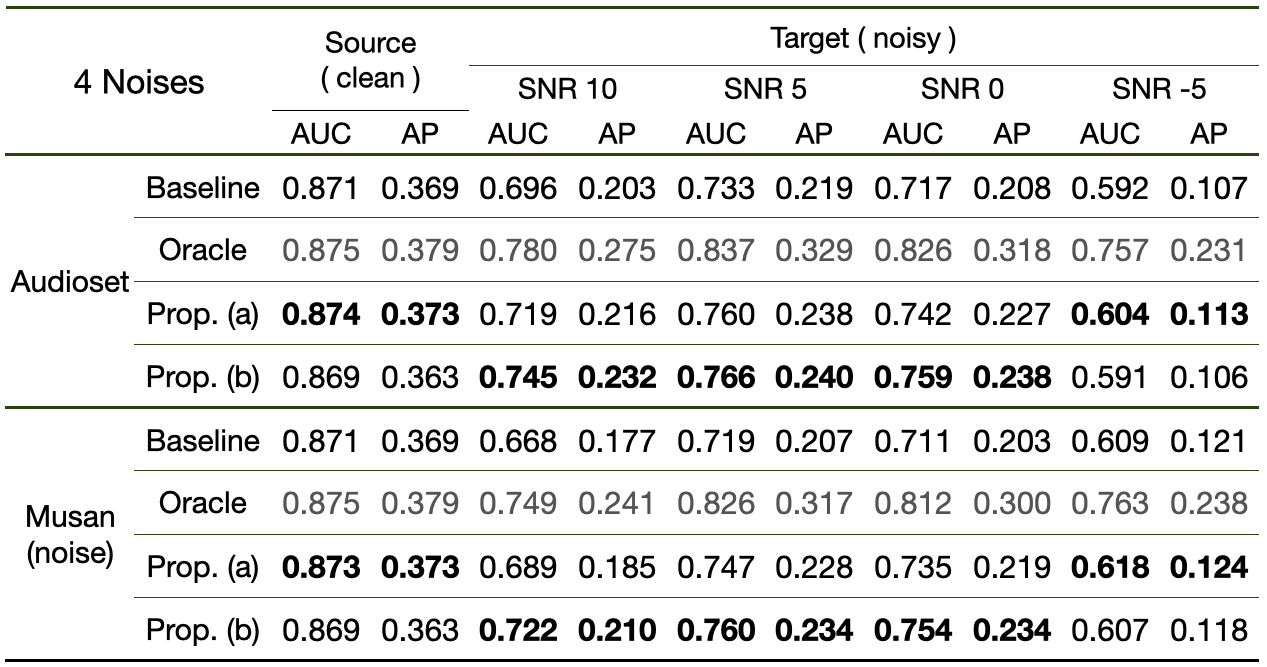}}}
    \caption{\centering The test  AUC and AP metrics with the inclusion of 4 noises.}
    \vspace{-0.4cm}
    \label{fig:noise1}
    \end{table}

\section{EXPERIMENT}
    
In the experiment, the Adam optimizer \cite{kingma2014adam} is used across all training phases. Specifically, the learning rate is set to 3e-4 for the pretraining of the FE, and 1e-4 for pretraining DC, finetuning FE, and training LP. The length of each audio input is fixed to 59,049 samples, in alignment with previous work \cite{spijkervet2021contrastive}. Furthermore, a uniform sample rate of 22,050 is applied across all training steps, with resampling if necessary.

To simulate various real-world noisy conditions, we synthesized music samples by combining them with one, two, or four different types of noise samples. During each data retrieval from the dataset, each music sample is normalized and randomly mixed with noise samples. The SNR for these mixtures is also randomly selected from a predefined range of [-10, 10]. For the validation and test phases, target samples are synthesized according to predefined SNRs as described in Figure 3.

For the evaluation metrics, we employed the area under the receiver operating characteristic curve (AUC) and average precision (AP). The experimental test results for source domain indicate that the baseline performance was either comparable to or slightly better than our proposed approach (Table 1, 2). This suggests that incorporating additional noisy music samples could potentially impact the existing performance adversely. However, as the number of noise samples increases, our proposed method demonstrates enhanced generalization and robustness (Table 3). Notably, the \textit{proposed (b)} configuration consistently delivered the best performance, underscoring the idea that even a modest amount of extra, noisy, and unlabeled data can improve the model's performance. Although the performance differences between \textit{proposed (a)} and \textit{proposed (b)} may appear subtle, it is important to note that the quantity of additional data samples per epoch in \textit{proposed (b)} is approximately 17 times less than the unlabeled data used in \textit{proposed (a)}.

Additionally, tests conducted on the Musan noise dataset corroborate the model's consistent performance across different noise conditions. These results further support the notion that the FE's embeddings are robust when exposed to a variety of noise types, thereby enhancing the model's overall generalization capabilities. Lastly, through all the evaluations, the model demonstrated enhanced performance even under challenging conditions involving various signal-to-noise ratios, affirming its robustness and utility for music auto-tagging in noisy scenarios.
\vspace{-0.2cm}

\section{CONCLUSION}

In this work, we introduced a novel framework for improving music auto-tagging performance by leveraging unlabeled noisy music data. We employed Domain Adversarial Training (DAT) to enhance the robustness of feature extraction, making it capable of handling both clean and noisy audio inputs effectively. Our experimental setup included diverse datasets such as MTG-Jamendo, MagnaTagATune (MTAT), Audioset, and Musan, and incorporated various real-world noise conditions to simulate realistic scenarios.

Our evaluations, using metrics such as AUC and AP, indicates promising results. While the incorporation of noisy data had a nuanced effect on performance, we found that as the variety of noise increased, the model's robustness improved, suggesting that our approach has strong generalization capabilities. In particular, the configuration using extra, unlabeled noisy data showed performance gains, even when the additional data volume was relatively small.

Additional tests on the Musan noise dataset corroborated the model's consistency and robustness across various noise conditions. These results affirm that the feature extractor's embeddings are resilient to noise, thereby extending the applicability and generalizability of our model for music auto-tagging in noisy environments.
\vspace{-0.4cm}

\section{Acknowledgement}
\vspace{-0.2cm}
This work was partly supported by Culture, Sports and Tourism R\&D Program through the Korea Creative Content Agency grant funded by the Ministry of Culture, Sports and Tourism in 2022 (No.R2022020066, 1/2) and the National Research Foundation of Korea (NRF) grant funded by the Korea government (MSIT) (No. RS-2023-00219429, 1/2).

\vfill\pagebreak

\bibliographystyle{IEEEbib}
\bibliography{Template}

\end{document}